\documentclass[12pt]{article}
\usepackage{a4wide}
\usepackage{amssymb}
\begin{document}
{\renewcommand{\thefootnote}{\fnsymbol{footnote}}
\begin{center}
{\LARGE  Effective field theory of loop quantum cosmology}\\
\vspace{1.5em}
Martin Bojowald\footnote{e-mail address: {\tt bojowald@gravity.psu.edu}}
\\
\vspace{0.5em}
Institute for Gravitation and the Cosmos,\\
The Pennsylvania State
University,\\
104 Davey Lab, University Park, PA 16802, USA\\
\vspace{1.5em}
\end{center}
}

\setcounter{footnote}{0}

\begin{abstract}
Quantum cosmology is traditionally formulated in a minisuperspace
  setting, implicitly averaging fields over space to obtain homogeneous
  models. For universal reasons related to the uncertainty principle, quantum
  corrections then depend on the size of the averaging volume. In
  minisuperspace truncations, the value of this volume remains an arbitrary
  parameter devoid of physical meaning, but in an effective field theory it
  is identified with the infrared scale of inhomogeneous modes. Moreover, the
  infared scale is running during gravitational collapse, when regions in
  which homogeneity remains approximately valid shrink to smaller and smaller
  co-moving sizes. Conceptual implications of this infrared renormalization
  for perturbative inhomogeneity in quantum cosmology are presented here,
  mainly for the example of loop quantum cosmology. Several claims made in
  this framework are altered by infrared renormalization.
\end{abstract}

\section{Introduction}

Quantum cosmology aims to perform two main tasks: to describe in a meaningful
way what is indicated by general relativity as the big-bang singularity, and
to derive potentially observable predictions in early-universe cosmology. Both
questions are of physical nature and therefore require a detailed
understanding of the relevant equations and solutions of Planck-scale
physics. In the absence of a complete theory of quantum gravity, effective
field theory provides important means to highlight implications that are
sufficiently generic to be insensitive to quantization ambiguities or choices
made in specific models. As we will see here, mainly in an application to loop
quantum cosmology \cite{LivRev,ROPP}, an effective field theory of quantum
cosmology indeed revises some claims extracted from a limited set of
simplified models.

These revisions can be split into two types: those related to the background
dynamics and those related to the behavior of inhomogeneity. As for the
former, a central role is played by the averaging volume $V_0$ used to express
an inhomogeneous geometry by a simplified homogeneous model. While the
classical equations are invariant under changing the size of this volume,
quantum corrections depend on this parameter as a consequence of uncertainty
relations: The averaging volume $V_0$ can be made arbitrarily small, such that
the geometrical volume $V=V_0a^3$ of a given region in an isotropic geometry
with scale factor $a$ has no lower bound. However, in a canonical
quantization, volume fluctuations $\Delta V$ are bounded from below by the
uncertainty relation $\Delta V\Delta H\geq \pi \ell_{\rm P}^2$ (noting that
$H/4\pi G$ is canonically conjugate to $V$). Since the Hubble parameter $H$
and the Planck length $\ell_{\rm P}$ are invariant under changing $V_0$, it is
impossible that fluctuations $(\Delta V,\Delta H)$ scale in the same way as
$(V,H)$ under changes of $V_0$. Quantum corrections from fluctuations
therefore generically break the classical $V_0$-invariance. (The fact that new
quantum effects may depend on the size of a region in which the theory is
formulated is well-known from the Casimir force.)

Canonical effective field theory, as reviewed in the next section, has
provided a physical interpretation for the appearance of $V_0$ in quantum
corrections of minisuperspace models \cite{MiniSup}: These corrections
describe the infrared contribution of quantum back-reaction in the underlying
inhomogeneous theory of which the minisuperspace model gives an averaged
description. The main new contribution of the present paper is an elaboration
of this result for perturbative inhomogeneity around the background of a
minisuperspace model. Because minisuperspace quantum corrections already
contain the infrared contribution from inhomogeneous modes, additional modes
brought in by perturbative inhomogeneity must be restricted to (co-moving)
wavelengths less than the cubic root of the averaging volume. Without this
restriction, some quantum corrections would be duplicated. (As another
consequence, it follows that $V_0$ is not an infrared regulator, which should
be sent to infinity after observables have been computed, but rather an
infrared scale which separates modes described in different ways --- by
averaging and more directly through perturbative inhomogeneity, respectively.)

If one uses a minisuperspace model amended by perturbative inhomogeneity to
probe deep quantum regimes, evolving toward increasing curvature, the
co-moving size of regions in which homogeneity is approximately realized
shrinks due to gravitational collapse. In order to maintain the approximation
implied by averaging, the value of $V_0$ should therefore be progressively
reduced. Combined with the interpretation of $V_0$ as an infrared scale,
reducing $V_0$ amounts to infrared renormalization. Based on the same
interpretation, the modes of perturbative inhomogeneity have to be restricted
to wavelengths less than the cubic root of $V_0$; therefore, the number of
modes changes during evolution, as a consequence of infrared renormalization
\cite{Infrared}. This behavior cannot be described by unitary evolution on a
single Hilbert space, but effective field theory provides the appropriate
setting.

Quantum cosmology has another effect on the behavior of inhomogeneity, related
to the form of quantum space-time structure.  In particular in the context of
theories such as loop quantum gravity that indicate some kind of spatial or
space-time discreteness, effective field theory is relevant also for
conceptual questions. For instance, while it is easy in minisuperspace models
to avoid divergences such as the big-bang singularity by modifying classical
equations in terms of bounded functions, this common practice raises several
questions: A modification in minisuperspace equations ignores consistency
conditions imposed by the requirement that equations for inhomogeneity be
covariant. If the theory resulting from such a modification is not consistent
with covariance, one has to demonstrate that one can still avoid low-energy
problems \cite{SmallLorentzViol}. Moreover, the physical nature of any
mechanism that helps to avoid the big-bang singularity often remains unclear
in minisuperspace models.  For instance, some studies in loop quantum
cosmology have claimed that the big bang is replaced by a ``bounce''
\cite{APS}. However, these models modify the gravitational terms in the
Friedmann equation while keeping the matter terms largely unchanged. How,
then, can it be possible that one evades singularity theorems without
violating energy conditions? These theorems do not use Einstein's equation or
the Friedmann equation, and should therefore remain valid after a modification
of the latter --- provided the modified theory has the same (Riemannian)
space-time structure as classical gravity, a property which is often assumed
implicitly in models of loop quantum cosmology
\cite{Hybrid,QFTCosmo,QFTCosmoClass,AAN}.

As we will discuss, a canonical effective field theory of quantum cosmology
provides insights into space-time structures and is thereby able to answer the
conceptual questions posed here. The two main topics of this paper ---
averaging and covariance --- are related by the presence of an infrared scale,
which is implied by averaging and has an effect on possible covariant
formulations.

\section{Averaging volume}

Going back to \cite{DeWitt}, minisuperspace models have traditionally been
used in quantum cosmology, in which one eliminates the spatial dependence of
all fields by averaging them over a specified region in space. The resulting
model has only temporal dependence, a fact which trivializes many consistency
conditions imposed by covariance. However, it turns out that minisuperspace
models remain sensitive to the covariance problem in a subtle way, which is
related to the coordinate volume of the region used to define the spatial
averaging.

This result has been illustrated in \cite{MiniSup} by a model of a
minisuperspace model, using a scalar field theory in flat space-time with
Lagrangian
\begin{equation} \label{L}
L=\int{\rm d}^3x
  \left(\frac{1}{2}\dot{\phi}^2-\frac{1}{2}|\nabla\phi|^2 -W(\phi)\right)\,.
\end{equation}
This theory has a minisuperspace model in which $\phi$ is assumed spatially
constant, such that the Lagrangian can be integrated over some region with
finite coordinate volume $V_0=\int{\rm d}^3x$:
\begin{equation}
 L_{\rm mini}= V_0 \left(\frac{1}{2}\dot{\phi}^2-W(\phi)\right)\,.
\end{equation}
We obtain the minisuperspace momentum
\begin{equation}
 p=\frac{\partial L_{\rm mini}}{\partial\dot{\phi}}=V_0\dot{\phi}
\end{equation}
and the Hamiltonian
\begin{equation}
 H_{\rm mini}= \frac{1}{2}\frac{p^2}{V_0}+V_0W(\phi)\,.
\end{equation}
The minisuperspace Hamiltonian can straightforwardly be quantized to
\begin{equation}
 \hat{H}_{\rm mini} = \frac{1}{2}\frac{\hat{p}^2}{V_0}+
   V_0W(\hat{\phi})\,. 
\end{equation}

In a semiclassical analysis, for instance using \cite{EffAc,Karpacz}, the
minisuperspace potential is found to have first-order corrections given by
\begin{equation} \label{Wmini}
 W^{\rm mini}_{\rm eff}(\phi)= W(\phi)+\frac{1}{2V_0} \hbar \sqrt{W''(\phi)}\,.
\end{equation}
The full scalar theory, on the other hand, has first-order corrections given by
the 
Coleman--Weinberg potential \cite{ColemanWeinberg}
\begin{equation}
 W_{\rm eff}(\phi) = W(\phi)+ \frac{1}{2}\hbar \int
 \frac{{\rm d}^4k}{(2\pi)^4} \log \left(1+
   \frac{W''(\phi)}{||{\bf k}||^2}\right)
\end{equation}
with an explicit $k^0$-integration \cite{CW}
\begin{equation}
 W_{\rm eff}(\phi) = W(\phi)+ \frac{1}{2}\hbar \int
 \frac{{\rm d}^3k}{(2\pi)^3}  \left(\sqrt{|\vec{k}|^2+W''(\phi)}-
   |\vec{k}|\right) \,.
\end{equation}
The minisuperspace potential (\ref{Wmini}) is therefore obtained as the
infrared contribution from quantum field theory: Integrating
\begin{equation} \label{WInfInt}
 \frac{1}{2}\hbar \int
 \frac{{\rm d}^3k}{(2\pi)^3}  \left(\sqrt{|\vec{k}|^2+W''(\phi)}-
   |\vec{k}|\right) 
\end{equation}
over $|\vec{k}|\leq k_{\rm max}= 2\pi/V_0^{1/3}$ results in
\begin{equation} \label{WInf}
 W_{\rm eff}(\phi) \approx W(\phi)+ \frac{\hbar}{12
  \pi^2} k_{\rm max}^3 \sqrt{W''(\phi)}= W(\phi)+\frac{2\pi}{3V_0}\hbar
\sqrt{W''(\phi)} \,.
\end{equation}

\subsection{Infrared renormalization}

We conclude that a minisuperspace truncation can capture some quantum effects
of the full theory, as long as $V_0\not\to\infty$.  If one sends $V_0$ to
infinity, as sometimes done following a misinterpretation of $V_0$ as an
infrared regulator \cite{InfReg}, all quantum corrections are erased. For
finite $V_0$, on the other hand, quantum corrections, unlike the classical
equations, depend on the seemingly arbitrary $V_0$. While minisuperspace
models could not explain the relevance of $V_0$, the connection with the
quantum-field theory (\ref{L}) makes it clear: $V_0$ represents the infrared
scale of modes included in a minisuperspace model through averaging. Quantum
effects of long-wavelength modes are therefore captured by minisuperspace
effective potentials. If short-wavelength modes are physically relevant, they
should be added onto the minisuperspace model, either as a full quantum-field
theory or as an effective field theory of perturbative inhomogeneity. In order
to avoid duplicating quantum corrections already contained in the infrared
contribution, perturbative inhomogeneity must be restricted to modes with
wavelengths less than $V_0^{1/3}$.

Based on (\ref{WInf}), a minisuperspace approximation is expected to be
reliable for large $V_0$ because it implicitly replaces the integration in
(\ref{WInfInt}) with the $k$-volume multiplied with the integrand at
$|\vec{k}|=0$. Geometrically, the averaging is justified only if there is no
strong inhomogeneity on scales less than $V_0$. This condition is realized at
late times in cosmology, thanks to large-scale homogeneity. However, if one
tries to use the same models to explain the big-bang singularity, the
Belinskii--Khalatnikov--Lifshitz (BKL) scenario \cite{BKL} shows that
generically one has to extend them to small $V_0$: This scenario indicates
that homogeneous dynamics may be relevant even close to a space-like
singularity, but only because the geometries at different spatial points
decouple from one another even while inhomogeneity grows on all scales. The
homogeneous dynamics indicated by BKL, viewed in a minisuperspace model, is
therefore compatible only with small $V_0$. If one evolves from small
curvature to large curvature, the averaging volume $V_0$ must be progressively
reduced in order to keep up with the shrinking scales of approximate
homogeneity. Adjusting the infrared scale $V_0$ amounts to infrared
renormalization \cite{Infrared}.

Infrared renormalization has two main implications. First, any quantum
treatment of perturbative inhomogeneity must be such that modes are restricted
to wavelenths less than $V_0^{1/3}$. As $V_0$ changes during infrared
renormalization, the number of field-theory modes changes as well. This
behavior cannot be modeled by unitary evolution on a fixed Hilbert space. At
this point, effective field theory is essential. Secondly, regarding the
background dynamics in minisuperspace models, small $V_0$ are relevant near
the big bang.  It turns out that models of loop quantum cosmology are
especially sensitive to this conclusion.

\subsection{Models of loop quantum gravity}

In isotropic models of loop quantum cosmology \cite{IsoCosmo}, one uses an
isotropic connection $A_a^i={c}\delta_a^i$ and a densitized triad
$E^b_j={p}\delta^b_j$ parameterized by a pair of canonical variables,
$c=\gamma\dot{a}$ and $|p|=a^2$. In this form, the flat-space classical
Friedmann equation is written as
\begin{equation}\label{Fried}
 -\frac{{c}^2}{\gamma^2|{p}|}+ \frac{8\pi G}{3}
 \rho=0\,.
\end{equation}

Loop quantization, however, does not provide an operator directly for $c$ but
only for matrix elements of holonomies \cite{LoopRep,ALMMT}, given in this
model by $\exp(i\mu c)$ with a real number $\mu$ \cite{Bohr}. In order to
become loop quantizable, the Friedmann equation is therefore modified by using
``holonomies,'' replacing ${c}^2/|{p}|$ with $\sin(\ell
{c}/\sqrt{|{p}|})^2/\ell^2$ in (\ref{Fried}), with some length parameter
$\ell$. (The $p$-dependence of the argument of the sine was first evaluated in
a cosmological setting in \cite{APSII} and is motivated by lattice refinement
\cite{InhomEvolve,CosConst}.)  Taken in isolation, holonomy modifications
indicate a ``bounce'' of isotropic models: The equation
\begin{equation}\label{ModFried}
 \frac{\sin(\ell {c}/\sqrt{|{p}|})^2}{\gamma^2\ell^2}= \frac{8\pi
   G}{3} \rho
\end{equation}
implies a bounded energy density which, for macroscopic $\rho$, remains true
in terms of expectation values in any nearly semiclassical state.

Gravitational effective actions are usually written in higher-curvature form;
see for instance \cite{EffectiveGR,BurgessLivRev}.  This well-known result is
a combination of two properties, the fact that quantum corrections generically
imply higher-derivative terms together with the condition of space-time
covariance. Covariance will be discussed in more detail in the next section,
and for now we focus on higher-derivative corrections or their canonical
analog.  The canonical approach to effective theory \cite{EffAc,Karpacz},
which is suitable for canonical quantum cosmology, realizes higher-derivative
corrections through auxiliary degrees of freedom, which have the useful
physical interpretation as moments of an evolving quantum state, back-reacting
on the trajectory of its basic expectation values. These terms therefore
originate from quantum-cosmological analogs of effective potentials such as
(\ref{Wmini}). As we just learned, these terms depend on the averaging volume
$V_0$, as can be seen also in explicit derivations such as
\cite{QuantumBounce,BounceSqueezed,FluctEn}.

An important consequence of the $V_0$-dependence can be seen easily if the
modified Friedmann equation (\ref{ModFried}) is rewritten as
\cite{AmbigConstr}
\begin{equation}\label{Hol}
 \frac{\dot{a}^2}{a^2} = \frac{8\pi G}{3} \rho\left(1-\frac{\rho}{\rho_{\rm
       QG}}\right)
\end{equation}
with
\begin{equation}
 \rho_{\rm QG}= \frac{3}{8\pi G\ell^2}\,.
\end{equation}
This equation is obtained by expressing the canonical momentum $c$ in terms of
$\dot{a}=\frac{1}{2}\dot{p}/\sqrt{|p|}$ using an equation of motion generated
by (\ref{ModFried}), and is therefore equivalent to (\ref{ModFried}).  

If quantum back-reaction is included, there is an additional coupling term to
quantum fluctuations and correlations which can be written as
\cite{QuantumBounce,BounceSqueezed}
\begin{equation}\label{Holsigma}
 \frac{\dot{a}^2}{a^2} = \frac{8\pi G}{3} \rho\left(1-\frac{\rho}{\rho_{\rm
       QG}}+\sigma\right)
\end{equation}
where 
\begin{equation} \label{sigma}
 \sigma= \frac{(\Delta V)^2-C+(2\pi G\ell\hbar/3)^2}{(V+2\pi G\ell\hbar/3)^2}
\end{equation}
and $V=V_0a^3$.  The fluctuation $\Delta V$ of $V$ appears explicitly in
(\ref{sigma}), and $C$ depends on the quantum correlation $C_{VH}$ between the
volume and its canonical momentum, the Hubble parameter $H$. Not much is known
about the quantum state of the universe, which makes it difficult to provide
an estimate of $C_{VH}$. However, for given fluctuations, quantum correlations
cannot be arbitrarily large owing to uncertainty relations. We may therefore
consider the term $C$ in (\ref{sigma}) less significant than $(\Delta V)$. The
expression (\ref{sigma}) can be used whenever higher-order moments of the
state are small compared with the second-order moments that appear in
(\ref{sigma}), that is, if the state is sufficiently semiclassical. More
generally, one can extend (\ref{sigma}) to a series in higher moments, as
given in \cite{QuantumBounce,BounceSqueezed}.

For $V\gg 2\pi G\ell\hbar/3$ (close to the Planck volume if $\ell\sim\ell_{\rm
  P}$) and a semiclassical state, we have $\sigma\ll 1$ and the $\sigma$-term
in (\ref{Holsigma}) can be ignored. However, when the term $\rho/\rho_{\rm
  QG}$ in (\ref{Holsigma}) is relevant, we expect to be close to a
classical singularity. We may then be justified in using homogeneous (although
anisotropic) models to analyze the local dynamics, appealing to the BKL
scenario. This scenario is asymptotic and does not set any lower bound, not
even the Planck volume, on the size $V_0$ of regions of approximate
homogeneity. Generically, we should therefore use small $V_0$ in this
regime. If $V\ll 2\pi G\ell\hbar/3$, even a semiclassical state results in
$\sigma\approx 1$, and quantum back-reaction cannot be ignored. Unfortunately,
small-$V_0$ solutions have so far been neglected in loop quantum
cosmology, following an influential claim \cite{APS} that argues for large
$V_0$ but overlooks infrared renormalization, the central lesson of the BKL
scenario for quantum cosmology.

The process of reducing the value of $V_0$ is related (but not equivalent) to
coarse-graining in which one views a continuum description as an averaged
microscopic formulation of many discrete patches. In coarse-graining, one is
led to consider small $V_0$ even in late-time cosmological eras in which one
is not forced to do so by inhomogeneity. In the context of loop quantum
cosmology, coarse graining has been studied, for instance, in
\cite{CoarseGrain}. In \cite{CoarseGrainSU11}, it has been observed that it is
possible to construct coherent states which, at least in some models, are
insensitive to coarse-graining in that their moments behave additively when
several such states are patched together. Such a behavior could be of interest
also in the context of changing $V_0$. However, even though the coherent state
in such a model is adapted to coarse-graining, it would still give rise to
$V_0$-dependent quantum corrections. In the example of \cite{CoarseGrainSU11},
for instance, the analog of the volume operator is called $\hat{z}$, and the
scale $V_0$ is replaced by a Casimir quantum number $j$. The expectation value
$\langle\hat{z}\rangle$ as well as the variance $(\Delta z)^2$ of $\hat{z}$
scale like $j$, such that a collection of $N$ coherent states behaves like a
single coherent state with scale $Nj$. However, the same proportionality
implies that $\langle\hat{z}^2\rangle=\langle\hat{z}\rangle^2+(\Delta z)^2$
does not have a clear scaling behavior since it is the sum of two terms, one
of which scales like $j^2$ and one like $j$. Therefore, quantum corrections
violate the classical scaling behavior of $z^2$ even in a coherent
state. (This violation can be traced back to generic properties of uncertainty
relations mentioned in the introduction.)

\section{Covariance}

Another set of quantum corrections in gravitational models is, generically,
given by higher-curvature contributions. In loop quantum cosmology, even if we
do not include higher-curvature corrections in the classical Friedmann
equation (\ref{Fried}), from which we set out to quantize a cosmological
model, such terms are expected in the semiclassical or effective dynamics that
describes an evolving wave function through the time dependence of its basic
expectation values. As explicitly shown in \cite{HigherTime}, such a dynamics
generically contains higher-derivative terms. If these terms are to descend
from a covariant action, they must be of higher-curvature type.
(In loop quantum gravity, the usual assumption is that covariance should not
be broken by quantum space-time effects, although it may be deformed. More
generally, it could be possible to construct theories that are not covariant
in the ultraviolet, as in \cite{Horava}, but have covariance restored in the
infrared; see for instance \cite{FoliationIR}. In such models, the
covariance-breaking terms in an action can usually be written as non-invariant
higher-curvature contribution, for instance using the extrinsic curvature of a
preferred foliation. Such theories are therefore included in the broad setting
of higher-curvature effective actions.)

Any effective dynamics of loop quantum cosmology, such as a complete version
of (\ref{ModFried}), should therefore contain isotropic reductions of
higher-curvature terms.  No such terms are included in (\ref{ModFried}), as
can easily be seen from the observation that they would imply
higher-derivative corrections (or auxiliary fields) that are absent in
(\ref{ModFried}). (This statement should not be confused with the existence of
higher-curvature analogs that can mimic the modified isotropic dynamics of
loop quantum cosmology \cite{ActionRhoSquared,LimCurvLQC,HigherDerivLQC} but
differ in the presence of anisotropies or perturbations
\cite{MimeticLQCPert,CovModPert,MimeticLQC}.)  Since models of loop quantum
cosmology are not based on a derivation from a covariant quantum theory,
suitable higher-curvature corrections have not been derived yet in this
framework. As long as they remain unknown, it is not justified to trust the
full function $\sin^2(\ell c/\sqrt{|p|})/\ell^2$ which, as a power series,
contains contributions of arbitrarily high order in $\ell^2c^2/|p|$. Instead,
one should use only the first term in the expansion, such as
\begin{equation}
 \frac{\sin(\ell
 {c}/\sqrt{|{p}|})^2}{\ell^2}\sim \frac{{c}^2}{|{p}|}\left(1-
 \frac{1}{3}\ell^2\frac{{c}^2}{|{p}|}+\cdots\right)\,.
\end{equation}
If $\ell\sim \ell_{\rm P}$, the leading corrections are $\ell_{\rm
  P}^2{c}^2/|{p}|\sim \rho/\rho_{\rm P}$, which is indeed of the same order as
expected for higher-curvature terms. Once suitable higher-curvature
corrections have been derived from loop quantum cosmology, they may well
change the high-density behavior; see for instance \cite{RegularizationLQC}.

\subsection{Problem of states}

Higher-derivative terms are a general feature in effective equations of
quantum mechanics, and they are state dependent. The low-energy limit of
effective actions often used in particle physics or for semiclassical gravity
\cite{BurgessLivRev,EffectiveGR} results in a unique effective action, but
only because an expansion around the ground state is assumed. In quantum
gravity it is not clear whether there is a suitable ground state, and even if
there is a candidate, it would likely not be the right choice for Planckian
physics near a spacelike singularity. 

This issue is exacerbated by the problem of time: Cosmology is dynamical and
therefore requires a reliable understanding of time. The classical equations
of cosmological models are time reparameterization invariant, which provides a
simplified setting in which covariance can be studied in quantum
cosmology. Moreover, the classical cosmological dynamics is constrained.

Consider a quantum constraint of the form
$\hat{C}=\hat{p}_{\phi}^2-\hat{H}^2$.  If $H$ does not depend on $\phi$, we
can write the constraint equation $\hat{C}\psi=0$ for states as an
``evolution'' equation
\begin{equation}
 -\hat{p}_{\phi}\psi=i\hbar\frac{\partial\psi}{\partial\phi}=\pm|\hat{H}|\psi
\end{equation}
However, the choice of $\phi$ as time affects quantum corrections
\cite{TwoTimes}, and it is unclear how one should choose relevant initial
states for $\phi$-``evolution.'' Moreover, in a cosmological model the
classical constraint is strongly restricted by its reduction from a covariant
theory. Consistency conditions are needed to find corresponding restrictions
on quantum corrections in a minisuperspace model, which does not have a clear
relationship with a covariant quantum theory.

Dealing with the problem of states requires parameterizations of a large class
of models, which again calls for an application of effective field theory. In
a canonical version, moments as auxiliary variables in an effective field
theory describe the freedom contained in the choice of states.  In the absence
of a distinguished state, quantum corrections, such as higher-derivative terms
from quantum back-reaction, depend on the choice of state.  As a consequence,
it is not clear whether the ``bounce'' claimed in loop quantum cosmology is
generic if higher-derivative terms are uncontrolled.  

\subsection{Space-time structure}

While the dynamical high-density behavior of models in loop quantum cosmology
remains ambiguous, the space-time structure at high density turns out to be
more controlled.  Canonical quantization does not presuppose the space-time
structure of a potential covariant quantum theory of gravity. Formulated in
terms of an action, both the Lagrangian density and the measure in
\begin{equation}
 S[g]= \frac{1}{16\pi G} \int{\rm d}^4x\:\sqrt{|\det g|}\:
 (R[g]+\cdots) 
\end{equation}
may then be subject to quantum corrections.  If this possibility is realized,
quantum-field theory on curved space-time, which presupposes a Riemannian
structure of space-time, is different from quantum gravity, in which the very
structure of space-time is likely modified by quantum effects.

Formal aspects of these statements may be illustrated for perturbative
inhomogeneity, assuming that we are perturbing the basic field as
$A(x)=\bar{A}+\delta A(x)$ which appears in a sample Hamiltonian
$h[A]:=A(x)^2$.  As in models of loop quantum gravity, a modified background
dynamics is then imposed by replacing $\bar{A}$ with $\ell^{-1}\sin(\ell
\bar{A})$.

The classical, unmodified theory has a perturbative Hamiltonian 
\begin{equation} 
h[\bar{A},\delta A] = \bar{A}^2+2\bar{A}\delta A(x)+
  \delta A(x)^2\,.
\end{equation}
A version of quantum-field theory on modified space-time, as proposed in
models of loop quantum cosmology in different guises
\cite{Hybrid,QFTCosmo,QFTCosmoClass,AAN}, has the perturbative Hamiltonian
\begin{equation}
h_{\ell}^{\rm QFT}[\bar{A},\delta A]=
  \ell^{-2}\sin(\ell\bar{A})^2+ 2\bar{A} \delta 
  A(x)+\delta A(x)^2
\end{equation}
in which classical equations are used for perturbations but not for the
background.  From an effective field theory of cosmological perturbations
within loop quantum cosmology
\cite{ConstraintAlgebra,ScalarHolInv,ScalarHolEv}, however, one obtains
\begin{equation}
h_{\ell}^{\rm QG}[\bar{A},\delta A]=
  \ell^{-2}\sin(\ell\bar{A})^2+ F_{\ell}(\bar{A}) \delta  
  A(x)+G_{\ell}(\bar{A})\delta A(x)^2
\end{equation}
with two functions $F_{\ell}$ and $G_{\ell}$ which are constrained by the
classical limit, $\lim_{\ell\to0}F_{\ell}(\bar{A})=2\bar{A}$ and
$\lim_{\ell\to0}G_{\ell}(\bar{A})=1$, as well as covariance conditions. The
latter are rather lengthy, but they imply that $F_{\ell}/\bar{A}$ and
$G_{\ell}$ are comparable to
$(\ell\bar{A})^{-2}\sin(\ell\bar{A})^2$. Therefore, they cannot be ignored if
the background dynamics is modified, presenting a crucial difference between
quantum-field theory on a modified background and effective quantum
cosmology. In particular, quantum-field theory on a modified background, as
used in \cite{Hybrid,QFTCosmo,QFTCosmoClass,AAN}, is inconsistent because it
ignores terms that are of the same order as crucial terms included in the
modification.

\subsection{Covariance from hypersurface deformations}

In canonical gravity, the generators $D[N^a]$ of deformations along a vector
field $N^a(x)$ tangential to a spatial slice together with the generators
$H[N]$ of normal deformations by a displacement of $N(x)$ obey brackets
\cite{DiracHamGR}
\begin{eqnarray}
 [D[N^a],D[M^b]]&=& -D[{\cal L}_{M^b}N^a] \label{DD}\\
{} [H[N],D[M^b]] &=& -H[{\cal L}_{M^b}N]\\
{} [H[N_1],H[N_2]] &=& D[q^{ab}(N_1\partial_bN_2-N_2\partial_bN_1)] \label{HH}
\end{eqnarray}
of a Lie algebroid \cite{ConsAlgebroid} in which the inverse of the induced
metric $q_{ab}$ on spatial slice appears in ``structure functions.''
Covariance in canonical quantum gravity then requires an anomaly-free
representation of these brackets by operators $\hat{D}$, $\hat{H}$, $\hat{q}$,
such that the commutators of smeared $\hat{D}$ and $\hat{H}$ are still
proportional to constraints and equal (\ref{DD})--(\ref{HH}) in a suitable
classical limit. 

Although there has been some recent progress
\cite{ThreeDeform,TwoPlusOneDef,TwoPlusOneDef2,AnoFreeWeak,AnoFreeWeakDiff,OffShell,ConstraintsG},
both the anomaly problem and the question of the classical limit remain open
in loop quantum gravity. It is, however, possible to analyze potential
outcomes by using methods of effective constraints \cite{EffCons,EffConsRel}
which, like effective Hamiltonians, are defined as expectation values of
constraint operators written as functions of basic expectation values and
moments. These variables are subject to a Poisson bracket which replaces the
commutator. In the classical limit, these effective constraints should
therefore obey the hypersurface-deformation brackets
 \begin{eqnarray}
 \{D[N^a],D[M^b]\}&=& -D[{\cal L}_{M^b}N^a]\\
 \{H[N],D[M^b]\} &=& -H[{\cal L}_{M^b}N]\\
 \{H[N_1],H[N_2]\} &=& D[q^{ab}(N_1\partial_bN_2-N_2\partial_bN_1)]
\end{eqnarray}
in Poisson-bracket form. 

It is important to note that this condition constitutes an off-shell property
that cannot be tested in gauge-fixed models. (See also \cite{NPZRev}.)  It turns
out that this condition is stronger than formal anomaly-freedom in a
reformulated system. For instance, it is possible to write the constraints in
spherically symmetric and polarized Gowdy models such that the
$\{H,H\}$-bracket with structure functions is replaced by an Abelian bracket
of the form $\{H+D,H+D\}=0$ \cite{LoopSchwarz,GowdyAbel}. However, these
formally consistent models are not always covariant in the sense just defined
\cite{SphSymmCov,GowdyCov}. Similarly, in some versions of the full theory it
is possible to rewrite the bracket with structure functions in the form
$\{H,H\}=\{D',D'\}$ and quantize this theory \cite{AnoFreeWeak} which,
however, does not imply that the covariance condition is maintained.

\subsection{Model}

Generic consequences of covariance in the presence of holonomy modifications
can be illustrated by another scalar model \cite{Loss}, now for a canonical
pair of a field $\phi(x)$ with momentum $p(x)$ in one spatial dimension.
The Hamiltonian and diffeomorphism constraints
\begin{equation}
 H[N]=\int{\rm d}x
 N\left(f(p)-\frac{1}{4}(\phi')^2-\frac{1}{2}\phi\phi''\right)\quad,\quad
 D[w]=\int{\rm d}x w\phi p' 
\end{equation}
can be chosen such that the latter generates
spatial diffeomorphisms,
\begin{equation}
 \delta_w\phi=\{\phi,D[w]\}= -(w\phi)' \quad,\quad
\delta_wp=\{p,D[w]\}=-wp'
\end{equation}
(showing that the field $\phi$ has density weight one), while the former has a
bracket
\begin{equation}
\{H[N],H[M]\}= D[\beta(p)(N'M-NM')]
\end{equation}
with $\beta(p)=\frac{1}{2} {\rm d}^2f/{\rm d}p^2$ that mimicks results from
detailed evaluations of covariance in models of loop quantum cosmology
\cite{ScalarHolInv,DeformedCosmo}. 

We have Lorentzian-type hypersurface deformations for $f(p)=p^2$.
With ``holonomy'' modifications $f(p)=p_0^2\sin^2(p/p_0)$, however, 
\begin{equation}
 \beta(p)=\frac{1}{2}{\rm d}^2f/{\rm d}p^2=\cos(2p/p_0)
\end{equation}
can be negative. In particular, at the maximum of $f(p)$,
\begin{equation}
\{H[N],H[M]\}= D[-(N'M-NM')]
\end{equation}
implies Euclidean signature: Linear $N$ and $M$ give boosts for $\beta(p)=1$,
such that $\Delta x=v\Delta t$, but rotations if $\beta(p)=-1$, such that
$\Delta x=-\theta \Delta y$ if $y$ is transversal to hypersurfaces.  The
opposite sign is also obtained if hypersurface-deformation brackets are
derived for Euclidean gravity, and using the methods of
\cite{Regained,LagrangianRegained} one can show that the brackets imply
elliptic field equations if $\beta(p)<0$ \cite{Action}.

In the scalar model, $\{H[N],D[w]\}$ does not close, but there are several
consistent gravity versions, including spherical symmetry and cosmological
perturbations \cite{JR,ScalarHolInv}.  The role of $p$ in the model is then
played by some curvature variable $K$, usually a component of extrinsic
curvature of space in the canonical splitting of space-time.  Replacing
$K^2\longrightarrow f(K)$ in the Hamiltonian constraint then modifies the
bracket such that
\begin{equation}
{} \{H[N_1],H[N_2]\} = D[\beta
q^{ab}(N_1\partial_bN_2-N_2\partial_bN_1)] 
\end{equation}
with 
\begin{equation}
 \beta(K)=\frac{1}{2} {\rm d}^2f(K)/{\rm d}K^2
=\cos(2\ell K)
\end{equation}
for $f(K)= \ell^{-2}\sin^2(\ell K)$, with a free parameter $\ell$ often
related to the Planck length.  For small $\ell K$, $\beta(K)\sim 1$, and the
classical hypersurface-deformation brackets are approximately realized. In
cosmological models, the small-$\ell K$ limit is usually achieved dynamically
at late times, but, more in the spirit of coarse-graining or renormalization,
one could interpret it as an infrared limit of the theory, akin to
\cite{FoliationIR}.

If $\ell K$ is not small, the structure of space-time can be subject to strong
modifications. In particular, we obtain signature change, $\beta(K)<0$, around
any local maximum of $f(K)$. In the same regime we would conclude, in a purely
homogeneous setting, that we obtain an upper bound of the energy density
(\ref{ModFried}). Since this point is surrounded by 4-dimensional
Euclidean-type space, however, any ``bounce'' obtained in this way is
indeterministic. (It is possible to have holonomy modifications without
signature change in models that use self-dual connections
\cite{SphSymmComplex,CosmoComplex,GowdyComplex} or implement only the
Euclidean version of the Hamiltonian constraint \cite{LQCScalar}. Given the
special form of simplified constraints in these models, the genericness of
this outcome remains unclear.)

These conclusions about modified structure functions are not undone by quantum
back-reaction or higher time derivatives \cite{EffConsQBR}. They are distinct
from higher-curvature corrections which would not alter the geometry of
hypersurface deformations \cite{HigherCurvHam}. In general, effects of
holonomy modifications cannot be described by an effective line element on a
standard space-time, as postulated in \cite{AAN}, because
${\rm d}x^a$ in
\begin{equation}
 {\rm d}s^2_{\rm
    eff}=\tilde{q}_{ab} {\rm d}x^a{\rm d}x^b
\end{equation}
do not transform by changes dual to deformed gauge transformations
$\{\tilde{q}_{ab},H[N]+D[w]\}$.  Field redefinitions to a standard $q_{ab}$
are possible as long as $\beta$ does not change sign \cite{Normal,EffLine}.
With signature change, however, we have a new model of non-classical
space-time. 

This result explains one of the questions posed in the
introduction: Bounce models of loop quantum cosmology can evade singularity
theorems even without violating energy conditions because they can be embedded
in inhomogeneous settings only with non-Riemannian space-time
structures. Since singularity theorems are formulated in the Riemannian
setting, there is no reason why they should be applicable. While this
observation resolves a conceptual question behind these bounce models, it also
sheds doubt on the usual deterministic interpretation of such a bounce: The
modified space-time structure implies signature change at high density, such
that the bounce regime is contained in a four-dimensional space without time
or a well-posed initial-value problem.

\subsection{Signature change}

A well-posed formulation of a mixed-type partial differential equation such as
\begin{equation} \label{Mixed}
 -\frac{\partial^2 u}{\partial t^2}+ \beta({\cal H}) \Delta u=0
\end{equation}
requires data on a characteristic where the equation is hyperbolic and on an
arc in the elliptic regime \cite{Tricomi}.  The arc implies that we need
future data for well-posedness, and therefore there is no deterministic
evolution. Mixed-type partial differential equations such as (\ref{Mixed}) are
more familiar from hydrodynamical descriptions of transonic flow, in which
case they give rise to the well-known sonic boom. Mathematically, this
phenomenon corresponds to a root-like pole, which generically appears in
solutions of (\ref{Mixed}) at the end of the elliptic arc \cite{Tricomi}. The
same consequence should be expected in cosmological applications
\cite{SigImpl}, implying a cosmic boom. While solutions $u$ of (\ref{Mixed})
are finite, their derivatives may diverge at isolated points. This behavior is
more well-behaved than the usual big-bang singularity, which implies diverging
derivatives of the space-time metric everywhere on a spatial
slice. Nevertheless, the existence of cosmic booms indicates that cosmological
perturbations are not sufficient for reliable solutions. The existence of
signature change, on the other hand, is more robust because it happens also in
spherically symmetric models with non-perturbative inhomogeneity
\cite{JR,HigherSpatial,SphSymmOp,MidiClass,BHSigChange,EffLine}. Signature
change in models of loop qantum cosmology means that instead of a ``bounce''
we have a non-singular beginning.

Signature change also has implications for black-hole models, and in
particular for the information-loss problem \cite{Loss}.  A non-singular,
non-rotating black-hole model can be envisioned by evolving through the
classical singularity using quantum evolution of the homogeneous interior. In
an inhomogenous completion, there is then no event horizon
\cite{BHInt,BHPara}.  However, consistent quantum space-time structure again
tells us that the high-curvature region is Euclidean.  A well-posed problem
requires arbitrary boundary values at the top boundary of the Euclidean
region, which affect the future space-time. In addition to the event horizon,
there is now a Cauchy horizon \cite{Loss}.

In contrast to these cautionary results, dynamical signature change from
holonomy modifications may be beneficial in a variety of scenarios. For
instance, non-commutative geometry \cite{ConnesBook} gives rise to discrete
spectra in Euclidean signature, which can help to regularize some of its
features \cite{NonCommQuanta}. Dynamical signature change is surprisingly
productive \cite{NoBoundLQC,LoopsRescue,NoBoundLQCGeo} in an application to
the no-boundary proposal \cite{nobound}. An analysis of the Lorentzian path
integral has recently revealed severe stability problems of perturbative
inhomogeneity with no-boundary initial conditions
\cite{LorentzianQC,NoSmooth,NoRescue}. Signature change from holonomy
modifications of loop quantum gravity changes the form of the relevant
off-shell instantons such that stability is found
\cite{LoopsRescue}. ``Off-shell'' here means that one does not impose the
Friedmann equation (or the Hamiltonian constraint) but only the equations of
motion it generates. The relationship (\ref{ModFried}) between holonomy
modifications and the energy density is therefore lost, and as a consequence
it is conceivable that signature change occurs at lower than Planckian
densities. This result is in fact borne out by a detailed analysis
\cite{LoopsRescue}, making signature change applicable at the sub-Planckian
densities used in the no-boundary proposal.

There is a final question which links the last result with our first
statements about the averaging volume. Even though off-shell instantons do not
impose the Friedmann equations, the background equations of motion they do use
should contain quantum corrections in which the averaging volume appears. In
contrast to our previous discussion of bounce models, infrared renormalization
now appears in a different form. In the no-boundary proposal, spatial slices
are assumed compact, such that they can foliate a 4-dimensional Euclidean
sphere rounding off space-time at the big bang. Near the pole of this 4-sphere
(replacing the big-bang singularity), the universe is nearly homogeneous
because it ``started'' in a single point, the pole, and inhomogeneous
perturbations are now under control thanks to the effects described in
\cite{LoopsRescue}. There is then no need for infrared renormalization, and
any constant $V_0$ may be used, such as the full coordinate volume $2\pi^2$ of
a unit 3-sphere. Therefore, $V_0$ does not become smaller and smaller as in
the BKL picture, and quantum corrections, just like perturbative
inhomogeneity, remain under control.

\section{Further directions}

In order to ensure reliable approximations, minisuperspace models can be
applied in the following way: Starting at low curvature, where our universe
indicates large-scale homogeneity and a large value of $V_0$ is consistent
with its role as an intrared scale, one progressively evolves toward larger
curvature. As the scale of homogeneity within a co-moving region is reduced by
gravitational collapse, the value of $V_0$ must be set smaller and smaller,
with implications for perturbative inhomogeneity as discussed in this paper. 

First, a fixed infrared scale means that modes of perturbative inhomogeneity
must be restricted to wave lengths less than $V_0^{1/3}$. Secondly, a fixed
infrared scale breaks covariance, but for consistency one must still ensure
that the infrared-fixed modes are obtained from a covariant theory. This
condition can be enforced by following the canonical procedure of implementing
the hypersurface-deformation algebroid, as derived for cosmological
perturbations in loop quantum cosmology in
\cite{ConstraintAlgebra,ScalarHolInv}. The resulting covariant equations
should then be evolved with a running infrared scale, a task which has not
been completed yet. The changing number of modes implies that effective field
theory is essential at this point and cannot be replaced by unitary evolution
on a fixed Hilbert space.

As $V_0$ becomes smaller and smaller, quantum corrections in the background
dynamics are magnified and, through their state dependence, the problem of
states gains in relevance. However, in such a regime the minisuperspace
approximation is much less justified because an infrared regulator is pushed
into the ultraviolet.  An effective field theory based on minisuperspace
approximations should break down at these scales, at the latest. It might, of
course, break down earlier, for instance in the presence of energy cascades as
in hydrodynamics: If energy is pumped into the system at large distance
scales, and dissipates due to friction at small distance scales, the distance
scales do not sufficiently decouple from one another to allow a reliable
effective evolution over long time scales. There may be a similar picture in
quantum cosmology, where expansion provides the analog of pumping energy into
inhomogeneous modes on large scales (particle production), while a modified
Planck-scale dynamics can often lead to new friction phenomena on small scales
\cite{Inflation} or repulsive contributions to the usually attractive
gravitational force \cite{LivRevOld}. The analysis of energy cascades in
quantum cosmology requires a deeper understanding of inhomogeneity, but even
if they are absent, it remains doubtful whether a minisuperspace treatment,
even one combined with perturbative inhomogeneity and infrared
renormalization, can reliably describe the fate of the big-bang singularity.
At least for the approach to high curvature a well-defined formulation is
available.

\section*{Acknowledgements}

This work was supported in part by NSF grant PHY-1607414.


\begin{thebibliography}{10}

\bibitem{LivRev}
M.\ Bojowald,
\newblock Loop Quantum Cosmology,
\newblock {\em Living Rev.\ Relativity} 11 (2008) 4, [gr-qc/0601085],
\newblock {\tt http://www.livingreviews.org/lrr-2008-4}

\bibitem{ROPP}
M.\ Bojowald,
\newblock Quantum cosmology: a review,
\newblock {\em Rep.\ Prog.\ Phys.} 78 (2015) 023901, [arXiv:1501.04899]

\bibitem{MiniSup}
M.\ Bojowald and S.\ Brahma,
\newblock Minisuperspace models as infrared contributions,
\newblock {\em Phys.\ Rev.\ D} 92 (2015) 065002, [arXiv:1509.00640]

\bibitem{Infrared}
M.\ Bojowald,
\newblock The BKL scenario, infrared renormalization, and quantum cosmology,
\newblock {\em JCAP} 01 (2019) 026, [arXiv:1810.00238]

\bibitem{SmallLorentzViol}
J.\ Polchinski,
\newblock Comment on [arXiv:1106.1417] ``Small Lorentz violations in quantum
  gravity: do they lead to unacceptably large effects?'', [arXiv:1106.6346]

\bibitem{APS}
A.\ Ashtekar, T.\ Pawlowski, and P.\ Singh,
\newblock Quantum Nature of the Big Bang: An Analytical and Numerical
  Investigation,
\newblock {\em Phys.\ Rev.\ D} 73 (2006) 124038, [gr-qc/0604013]

\bibitem{Hybrid}
M.\ Mart{\'\i}n-Benito, L.~J.\ Garay, and G.~A.\ Mena~Marug\'an,
\newblock Hybrid Quantum Gowdy Cosmology: Combining Loop and Fock
  Quantizations,
\newblock {\em Phys.\ Rev.\ D} 78 (2008) 083516, [arXiv:0804.1098]

\bibitem{QFTCosmo}
A.\ Ashtekar, W.\ Kaminski, and J.\ Lewandowski,
\newblock Quantum field theory on a cosmological, quantum space-time,
\newblock {\em Phys.\ Rev.\ D} 79 (2009) 064030, [arXiv:0901.0933]

\bibitem{QFTCosmoClass}
A.\ Dapor, J.\ Lewandowski, and J.\ Puchta,
\newblock QFT on quantum spacetime: a compatible classical framework,
\newblock {\em Phys.\ Rev.\ D} 87 (2013) 104038, [arXiv:1302.3038]

\bibitem{AAN}
I.\ Agull\'o, A.\ Ashtekar, and W.\ Nelson,
\newblock An Extension of the Quantum Theory of Cosmological Perturbations to
  the Planck Era,
\newblock {\em Phys.\ Rev.\ D} 87 (2013) 043507, [arXiv:1211.1354]

\bibitem{DeWitt}
B.~S.\ DeWitt,
\newblock Quantum Theory of Gravity. I. The Canonical Theory,
\newblock {\em Phys.\ Rev.} 160 (1967) 1113--1148

\bibitem{EffAc}
M.\ Bojowald and A.\ Skirzewski,
\newblock Effective Equations of Motion for Quantum Systems,
\newblock {\em Rev.\ Math.\ Phys.} 18 (2006) 713--745, [math-ph/0511043]

\bibitem{Karpacz}
M.\ Bojowald and A.\ Skirzewski,
\newblock Quantum Gravity and Higher Curvature Actions,
\newblock {\em Int.\ J.\ Geom.\ Meth.\ Mod.\ Phys.} 4 (2007) 25--52,
  [hep-th/0606232],
\newblock Proceedings of ``Current Mathematical Topics in Gravitation and
  Cosmology'' (42nd Karpacz Winter School of Theoretical Physics), Ed.\
  Borowiec, A.\ and Francaviglia, M.

\bibitem{ColemanWeinberg}
S.\ Coleman and E.\ Weinberg,
\newblock Radiative corrections as the origin of spontaneous symmetry breaking,
\newblock {\em Phys.\ Rev.\ D} 7 (1973) 1888--1910

\bibitem{CW}
M.\ Bojowald and S.\ Brahma,
\newblock Canonical derivation of effective potentials
\newblock (2014), [arXiv:1411.3636]

\bibitem{InfReg}
C.\ Rovelli and E.\ Wilson-Ewing,
\newblock Why are the effective equations of loop quantum cosmology so
  accurate?,
\newblock {\em Phys.\ Rev.\ D} 90 (2014) 023538, [arXiv:1310.8654]

\bibitem{BKL}
V.~A.\ Belinskii, I.~M.\ Khalatnikov, and E.~M.\ Lifschitz,
\newblock A general solution of the Einstein equations with a time singularity,
\newblock {\em Adv.\ Phys.} 13 (1982) 639--667

\bibitem{IsoCosmo}
M.\ Bojowald,
\newblock Isotropic Loop Quantum Cosmology,
\newblock {\em Class.\ Quantum Grav.} 19 (2002) 2717--2741, [gr-qc/0202077]

\bibitem{LoopRep}
C.\ Rovelli and L.\ Smolin,
\newblock Loop Space Representation of Quantum General Relativity,
\newblock {\em Nucl.\ Phys.\ B} 331 (1990) 80--152

\bibitem{ALMMT}
A.\ Ashtekar, J.\ Lewandowski, D.\ Marolf, J.\ Mour\~ao, and T.\ Thiemann,
\newblock Quantization of Diffeomorphism Invariant Theories of Connections with
  Local Degrees of Freedom,
\newblock {\em J.\ Math.\ Phys.} 36 (1995) 6456--6493, [gr-qc/9504018]

\bibitem{Bohr}
A.\ Ashtekar, M.\ Bojowald, and J.\ Lewandowski,
\newblock Mathematical structure of loop quantum cosmology,
\newblock {\em Adv.\ Theor.\ Math.\ Phys.} 7 (2003) 233--268, [gr-qc/0304074]

\bibitem{APSII}
A.\ Ashtekar, T.\ Pawlowski, and P.\ Singh,
\newblock Quantum Nature of the Big Bang: Improved dynamics,
\newblock {\em Phys.\ Rev.\ D} 74 (2006) 084003, [gr-qc/0607039]

\bibitem{InhomEvolve}
M.\ Bojowald, H.\ Hern\'andez, M.\ Kagan, P.\ Singh, and A.\ Skirzewski,
\newblock Formation and evolution of structure in loop cosmology,
\newblock {\em Phys.\ Rev.\ Lett.} 98 (2007) 031301, [astro-ph/0611685]

\bibitem{CosConst}
M.\ Bojowald,
\newblock The dark side of a patchwork universe,
\newblock {\em Gen.\ Rel.\ Grav.} 40 (2008) 639--660, [arXiv:0705.4398]

\bibitem{EffectiveGR}
J.~F.\ Donoghue,
\newblock General relativity as an effective field theory: The leading quantum
  corrections,
\newblock {\em Phys.\ Rev.\ D} 50 (1994) 3874--3888, [gr-qc/9405057]

\bibitem{BurgessLivRev}
C.~P.\ Burgess,
\newblock Quantum Gravity in Everyday Life: General Relativity as an Effective
  Field Theory,
\newblock {\em Living Rev.\ Relativity} 7 (2004), [gr-qc/0311082],
\newblock http://www.livingreviews.org/lrr-2004-5

\bibitem{QuantumBounce}
M.\ Bojowald,
\newblock Quantum nature of cosmological bounces,
\newblock {\em Gen.\ Rel.\ Grav.} 40 (2008) 2659--2683, [arXiv:0801.4001]

\bibitem{BounceSqueezed}
M.\ Bojowald,
\newblock How quantum is the big bang?,
\newblock {\em Phys.\ Rev.\ Lett.} 100 (2008) 221301, [arXiv:0805.1192]

\bibitem{FluctEn}
M.\ Bojowald,
\newblock Fluctuation energies in quantum cosmology,
\newblock {\em Phys.\ Rev.\ D} 89 (2014) 124031, [arXiv:1404.5284]

\bibitem{AmbigConstr}
K.\ Vandersloot,
\newblock On the Hamiltonian Constraint of Loop Quantum Cosmology,
\newblock {\em Phys.\ Rev.\ D} 71 (2005) 103506, [gr-qc/0502082]

\bibitem{CoarseGrain}
N.\ Bodendorfer,
\newblock State refinements and coarse graining in a full theory embedding of
  loop quantum cosmology,
\newblock {\em Class.\ Quantum Grav.} 34 (2017) 135016, [arXiv:1607.06227]

\bibitem{CoarseGrainSU11}
N.\ Bodendorfer and F.\ Haneder,
\newblock Coarse graining as a representation change,
\newblock {\em Phys.\ Lett.\ B} 792 (2019) 69--73, [arXiv:1811.02792]

\bibitem{HigherTime}
M.\ Bojowald, S.\ Brahma, and E.\ Nelson,
\newblock Higher time derivatives in effective equations of canonical quantum
  systems,
\newblock {\em Phys.\ Rev.\ D} 86 (2012) 105004, [arXiv:1208.1242]

\bibitem{Horava}
P.\ Ho\v{r}ava,
\newblock Quantum gravity at a Lifshitz point,
\newblock {\em Phys.\ Rev.\ D} 79 (2009) 084008, [arXiv:0901.3775]

\bibitem{FoliationIR}
T.\ Koslowski and A.\ Schenkel,
\newblock Preferred foliation effects in Quantum General Relativity,
\newblock {\em Class.\ Quant.\ Grav.} 27 (2010) 135014, [arXiv:0910.0623]

\bibitem{ActionRhoSquared}
G.~J.\ Olmo and P.\ Singh,
\newblock Covariant Effective Action for Loop Quantum Cosmology a la Palatini,
\newblock {\em JCAP} 0901 (2009) 030, [arXiv:0806.2783]

\bibitem{LimCurvLQC}
N.\ Bodendorfer, A.\ Sch\"afer, and J.\ Schliemann,
\newblock On the canonical structure of general relativity with a limiting
  curvature and its relation to loop quantum gravity,
\newblock {\em Phys.\ Rev.\ D} 97 (2018) 084057, [arXiv:1703.10670]

\bibitem{HigherDerivLQC}
D.\ Langlois, H.\ Liu, K.\ Noui, and E.\ Wilson-Ewing,
\newblock Effective loop quantum cosmology as a higher-derivative scalar-tensor
  theory,
\newblock {\em Class.\ Quant.\ Grav.} 34 (2017) 225004, [arXiv:1703.10812]

\bibitem{MimeticLQCPert}
J.\ Haro, Ll.\ Arest\'e~Sal\'o, and S.\ Pan,
\newblock Mimetic Loop Quantum Cosmology, [arXiv:1803.09653]

\bibitem{CovModPert}
J.\ Haro, Ll.\ Arest\'e~Sal\'o, and E.\ Elizalde,
\newblock Cosmological perturbations in a class of fully covariant modified
  theories: Application to models with the same background as standard LQC,
\newblock {\em Eur.\ Phys.\ J.\ C} 78 (2018) 712, [arXiv:1806.07196]

\bibitem{MimeticLQC}
N.\ Bodendorfer, F.~M.\ Mele, and J.\ M\"unch,
\newblock Is limiting curvature mimetic gravity an effective polymer quantum
  gravity?,
\newblock {\em Class.\ Quantum Grav.} 35 (2018) 225001, [arXiv:1806.02052]

\bibitem{RegularizationLQC}
R.\ Helling,
\newblock Higher curvature counter terms cause the bounce in loop cosmology,
  [arXiv:0912.3011]

\bibitem{TwoTimes}
M.\ Bojowald and T.\ Halnon,
\newblock Time in quantum cosmology,
\newblock {\em Phys.\ Rev.\ D} 98 (2018) 066001, [arXiv:1612.00353]

\bibitem{ConstraintAlgebra}
M.\ Bojowald, G.\ Hossain, M.\ Kagan, and S.\ Shankaranarayanan,
\newblock Anomaly freedom in perturbative loop quantum gravity,
\newblock {\em Phys.\ Rev.\ D} 78 (2008) 063547, [arXiv:0806.3929]

\bibitem{ScalarHolInv}
T.\ Cailleteau, L.\ Linsefors, and A.\ Barrau,
\newblock Anomaly-free perturbations with inverse-volume and holonomy
  corrections in Loop Quantum Cosmology,
\newblock {\em Class.\ Quantum Grav.} 31 (2014) 125011, [arXiv:1307.5238]

\bibitem{ScalarHolEv}
E.\ Wilson-Ewing,
\newblock Holonomy Corrections in the Effective Equations for Scalar Mode
  Perturbations in Loop Quantum Cosmology,
\newblock {\em Class.\ Quant.\ Grav.} 29 (2012) 085005, [arXiv:1108.6265]

\bibitem{DiracHamGR}
P.~A.~M.\ Dirac,
\newblock The theory of gravitation in Hamiltonian form,
\newblock {\em Proc.\ Roy.\ Soc.\ A} 246 (1958) 333--343

\bibitem{ConsAlgebroid}
C.\ Blohmann, M.~C.\ Barbosa~Fernandes, and A.\ Weinstein,
\newblock Groupoid symmetry and constraints in general relativity. 1:
  kinematics,
\newblock {\em Commun.\ Contemp.\ Math.} 15 (2013) 1250061, [arXiv:1003.2857]

\bibitem{ThreeDeform}
A.\ Perez and D.\ Pranzetti,
\newblock On the regularization of the constraints algebra of Quantum Gravity
  in $2+1$ dimensions with non-vanishing cosmological constant,
\newblock {\em Class.\ Quantum Grav.} 27 (2010) 145009, [arXiv:1001.3292]

\bibitem{TwoPlusOneDef}
A.\ Henderson, A.\ Laddha, and C.\ Tomlin,
\newblock Constraint algebra in LQG reloaded : Toy model of a ${\rm U}(1)^{3}$
  Gauge Theory I,
\newblock {\em Phys.\ Rev.\ D} 88 (2013) 044028, [arXiv:1204.0211]

\bibitem{TwoPlusOneDef2}
A.\ Henderson, A.\ Laddha, and C.\ Tomlin,
\newblock Constraint algebra in LQG reloaded : Toy model of an Abelian gauge
  theory -- II Spatial Diffeomorphisms,
\newblock {\em Phys.\ Rev.\ D} 88 (2013) 044029, [arXiv:1210.3960]

\bibitem{AnoFreeWeak}
C.\ Tomlin and M.\ Varadarajan,
\newblock Towards an Anomaly-Free Quantum Dynamics for a Weak Coupling Limit of
  Euclidean Gravity,
\newblock {\em Phys.\ Rev.\ D} 87 (2013) 044039, [arXiv:1210.6869]

\bibitem{AnoFreeWeakDiff}
M.\ Varadarajan,
\newblock Towards an Anomaly-Free Quantum Dynamics for a Weak Coupling Limit of
  Euclidean Gravity: Diffeomorphism Covariance,
\newblock {\em Phys.\ Rev.\ D} 87 (2013) 044040, [arXiv:1210.6877]

\bibitem{OffShell}
A.\ Laddha,
\newblock Hamiltonian constraint in Euclidean LQG revisited: First hints of
  off-shell Closure
\newblock (2014), [arXiv:1401.0931]

\bibitem{ConstraintsG}
M.\ Varadarajan,
\newblock The constraint algebra in Smolins' $G\to 0$ limit of 4d Euclidean
  Gravity,
\newblock {\em Phys.\ Rev.\ D} 97 (2018) 106007, [arXiv:1802.07033]

\bibitem{EffCons}
M.\ Bojowald, B.\ Sandh\"ofer, A.\ Skirzewski, and A.\ Tsobanjan,
\newblock Effective constraints for quantum systems,
\newblock {\em Rev.\ Math.\ Phys.} 21 (2009) 111--154, [arXiv:0804.3365]

\bibitem{EffConsRel}
M.\ Bojowald and A.\ Tsobanjan,
\newblock Effective constraints for relativistic quantum systems,
\newblock {\em Phys.\ Rev.\ D} 80 (2009) 125008, [arXiv:0906.1772]

\bibitem{NPZRev}
H.\ Nicolai, K.\ Peeters, and M.\ Zamaklar,
\newblock Loop quantum gravity: an outside view,
\newblock {\em Class.\ Quantum Grav.} 22 (2005) R193--R247, [hep-th/0501114]

\bibitem{LoopSchwarz}
R.\ Gambini and J.\ Pullin,
\newblock Loop quantization of the Schwarzschild black hole,
\newblock {\em Phys.\ Rev.\ Lett.} 110 (2013) 211301, [arXiv:1302.5265]

\bibitem{GowdyAbel}
D.\ Mart\'{\i}n-de Blas, J.\ Olmedo, and T.\ Pawlowski,
\newblock Loop quantization of the Gowdy model with local rotational symmetry,
  [arXiv:1509.09197]

\bibitem{SphSymmCov}
M.\ Bojowald, S.\ Brahma, and J.~D.\ Reyes,
\newblock Covariance in models of loop quantum gravity: Spherical symmetry,
\newblock {\em Phys.\ Rev.\ D} 92 (2015) 045043, [arXiv:1507.00329]

\bibitem{GowdyCov}
M.\ Bojowald and S.\ Brahma,
\newblock Covariance in models of loop quantum gravity: Gowdy systems,
\newblock {\em Phys.\ Rev.\ D} 92 (2015) 065002, [arXiv:1507.00679]

\bibitem{Loss}
M.\ Bojowald,
\newblock Information loss, made worse by quantum gravity,
\newblock {\em Front.\ Phys.} 3 (2015) 33, [arXiv:1409.3157]

\bibitem{DeformedCosmo}
A.\ Barrau, M.\ Bojowald, G.\ Calcagni, J.\ Grain, and M.\ Kagan,
\newblock Anomaly-free cosmological perturbations in effective canonical
  quantum gravity,
\newblock {\em JCAP} 05 (2015) 051, [arXiv:1404.1018]

\bibitem{Regained}
S.~A.\ Hojman, K.\ Kucha\v{r}, and C.\ Teitelboim,
\newblock Geometrodynamics Regained,
\newblock {\em Ann.\ Phys.\ (New York)} 96 (1976) 88--135

\bibitem{LagrangianRegained}
K.~V.\ Kucha\v{r},
\newblock Geometrodynamics regained: A Lagrangian approach,
\newblock {\em J.\ Math.\ Phys.} 15 (1974) 708--715

\bibitem{Action}
M.\ Bojowald and G.~M.\ Paily,
\newblock Deformed General Relativity and Effective Actions from Loop Quantum
  Gravity,
\newblock {\em Phys.\ Rev.\ D} 86 (2012) 104018, [arXiv:1112.1899]

\bibitem{JR}
J.~D.\ Reyes,
\newblock {\em Spherically Symmetric Loop Quantum Gravity: Connections to
  2-Dimensional Models and Applications to Gravitational Collapse},
\newblock PhD thesis, The Pennsylvania State University, 2009

\bibitem{SphSymmComplex}
J.\ Ben~Achour, S.\ Brahma, and A.\ Marciano,
\newblock Spherically symmetric sector of self dual Ashtekar gravity coupled to
  matter: Anomaly-free algebra of constraints with holonomy corrections,
\newblock {\em Phys.\ Rev.\ D} 96 (2017) 026002, [arXiv:1608.07314]

\bibitem{CosmoComplex}
J.\ Ben~Achour, S.\ Brahma, J.\ Grain, and A.\ Marciano,
\newblock A new look at scalar perturbations in loop quantum cosmology:
  (un)deformed algebra approach using self dual variables
\newblock (2016), [arXiv:1610.07467]

\bibitem{GowdyComplex}
J.\ Ben~Achour and S.\ Brahma,
\newblock Covariance in self dual inhomogeneous models of effective quantum
  geometry: Spherical symmetry and Gowdy systems,
\newblock {\em Phys.\ Rev.\ D} 97 (2018) 126003, [arXiv:1712.03677]

\bibitem{LQCScalar}
J.-P.\ Wu, M.\ Bojowald, and Y.\ Ma,
\newblock Anomaly freedom in perturbative models of Euclidean loop quantum
  gravity,
\newblock {\em Phys.\ Rev.\ D} 98 (2018) 106009, [arXiv:1809.04465]

\bibitem{EffConsQBR}
M.\ Bojowald and S.\ Brahma,
\newblock Effective constraint algebras with structure functions,
\newblock {\em J. Phys. A: Math. Theor.} 49 (2016) 125301, [arXiv:1407.4444]

\bibitem{HigherCurvHam}
N.\ Deruelle, M.\ Sasaki, Y.\ Sendouda, and D.\ Yamauchi,
\newblock Hamiltonian formulation of $f({\rm Riemann})$ theories of gravity,
\newblock {\em Prog.\ Theor.\ Phys.} 123 (2010) 169--185, [arXiv:0908.0679]

\bibitem{Normal}
M.\ Bojowald, S.\ Brahma, U.\ B\"{u}y\"{u}k\c{c}am, and F.\ D'Ambrosio,
\newblock Hypersurface-deformation algebroids and effective space-time models,
\newblock {\em Phys.\ Rev.\ D} 94 (2016) 104032, [arXiv:1610.08355]

\bibitem{EffLine}
M.\ Bojowald, S.\ Brahma, and D.-H.\ Yeom,
\newblock Effective line elements and black-hole models in canonical (loop)
  quantum gravity,
\newblock {\em Phys.\ Rev.\ D} 98 (2018) 046015, [arXiv:1803.01119]

\bibitem{Tricomi}
F.~G.\ Tricomi,
\newblock {\em Repertorium der Theorie der Differentialgleichungen},
\newblock Springer Verlag, 1968

\bibitem{SigImpl}
M.\ Bojowald and J.\ Mielczarek,
\newblock Some implications of signature-change in cosmological models of loop
  quantum gravity,
\newblock {\em JCAP} 08 (2015) 052, [arXiv:1503.09154]

\bibitem{HigherSpatial}
M.\ Bojowald, G.~M.\ Paily, and J.~D.\ Reyes,
\newblock Discreteness corrections and higher spatial derivatives in effective
  canonical quantum gravity,
\newblock {\em Phys.\ Rev.\ D} 90 (2014) 025025, [arXiv:1402.5130]

\bibitem{SphSymmOp}
S.\ Brahma,
\newblock Spherically symmetric canonical quantum gravity,
\newblock {\em Phys.\ Rev.\ D} 91 (2015) 124003, [arXiv:1411.3661]

\bibitem{MidiClass}
M.\ Bojowald and S.\ Brahma,
\newblock Signature change in loop quantum gravity: Two-dimensional
  midisuperspace models and dilaton gravity,
\newblock {\em Phys.\ Rev.\ D} 95 (2017) 124014, [arXiv:1610.08840]

\bibitem{BHSigChange}
M.\ Bojowald and S.\ Brahma,
\newblock Signature change in 2-dimensional black-hole models of loop quantum
  gravity,
\newblock {\em Phys.\ Rev.\ D} 98 (2018) 026012, [arXiv:1610.08850]

\bibitem{BHInt}
A.\ Ashtekar and M.\ Bojowald,
\newblock Quantum Geometry and the Schwarzschild Singularity,
\newblock {\em Class.\ Quantum Grav.} 23 (2006) 391--411, [gr-qc/0509075]

\bibitem{BHPara}
A.\ Ashtekar and M.\ Bojowald,
\newblock Black hole evaporation: A paradigm,
\newblock {\em Class.\ Quantum Grav.} 22 (2005) 3349--3362, [gr-qc/0504029]

\bibitem{ConnesBook}
A.\ Connes,
\newblock {\em Non-commutative geometry},
\newblock Academic Press, Boston, MA, 1994

\bibitem{NonCommQuanta}
A.~H.\ Chamseddine, A.\ Connes, and V.\ Mukhanov,
\newblock Quanta of Geometry: Noncommutative Aspects,
\newblock {\em Phys.\ Rev.\ Lett.} 114 (2015) 091302, [arXiv:1409.2471]

\bibitem{NoBoundLQC}
S.\ Brahma and D.-h.\ Yeom,
\newblock The no-boundary wave function for loop quantum cosmology,
\newblock {\em Phys.\ Rev.\ D} 98 (2018) 083537, [arXiv:1808.01744]

\bibitem{LoopsRescue}
M.\ Bojowald and S.\ Brahma,
\newblock Loops rescue the no-boundary proposal,
\newblock {\em Phys.\ Rev.\ Lett.} 121 (2018) 201301, [arXiv:1810.09871]

\bibitem{NoBoundLQCGeo}
S.\ Brahma and D.-h.\ Yeom,
\newblock On the geometry of no-boundary instantons in loop quantum cosmology,
  [arXiv:1810.10211]

\bibitem{nobound}
J.~B.\ Hartle and S.~W.\ Hawking,
\newblock Wave function of the Universe,
\newblock {\em Phys.\ Rev.\ D} 28 (1983) 2960--2975

\bibitem{LorentzianQC}
J.\ Feldbrugge, J.-L.\ Lehners, and N.\ Turok,
\newblock Lorentzian Quantum Cosmology,
\newblock {\em Phys.\ Rev.\ D} 95 (2017) 103508, [arXiv:1703.02076]

\bibitem{NoSmooth}
J.\ Feldbrugge, J.-L.\ Lehners, and N.\ Turok,
\newblock No smooth beginning for spacetime,
\newblock {\em Phys.\ Rev.\ Lett.} 119 (2017) 171301, [arXiv:1705.00192]

\bibitem{NoRescue}
J.\ Feldbrugge, J.-L.\ Lehners, and N.\ Turok,
\newblock No rescue for the no boundary proposal: Pointers to the future of
  quantum cosmology,
\newblock {\em Phys.\ Rev.\ D} 97 (2018) 023509, [arXiv:1708.05104]

\bibitem{Inflation}
M.\ Bojowald,
\newblock Inflation from quantum geometry,
\newblock {\em Phys.\ Rev.\ Lett.} 89 (2002) 261301, [gr-qc/0206054]

\bibitem{LivRevOld}
M.\ Bojowald,
\newblock Loop Quantum Cosmology,
\newblock {\em Living Rev.\ Relativity} 8 (2005) 11, [gr-qc/0601085],
\newblock {\tt http://www.livingreviews.org/lrr-2005-11}

\end{thebibliography}

\end{document}